\documentclass[11pt]{article}

\usepackage[margin=1in]{geometry}
\usepackage[T1]{fontenc}
\usepackage[utf8]{inputenc}
\usepackage{lmodern}
\usepackage{microtype}
\usepackage{setspace}

\usepackage{authblk}

\usepackage{amsmath,amssymb,bm}
\usepackage{siunitx}
\usepackage{graphicx}
\usepackage{booktabs}
\usepackage{array}
\usepackage{tabularx}
\usepackage{longtable}
\usepackage{adjustbox}
\usepackage{multirow}
\usepackage{makecell}
\usepackage{float}
\usepackage{placeins}
\usepackage{caption}
\usepackage{subcaption}

\newcolumntype{Y}{>{\raggedright\arraybackslash}X}

\usepackage{enumitem}
\setlist[itemize]{leftmargin=1.5em}
\setlist[enumerate]{leftmargin=1.5em}
\usepackage{xcolor}
\usepackage{xurl}
\usepackage[numbers,sort&compress]{natbib}
\usepackage{hyperref}
\usepackage[nameinlink,noabbrev]{cleveref}

\hypersetup{
    colorlinks=true,
    linkcolor=blue!50!black,
    citecolor=blue!50!black,
    urlcolor=blue!50!black,
    pdftitle={Evolutionary Period-Change Modeling of Delta Cephei with MESA Tracks},
    pdfauthor={Zuhoor Elahi, Christopher Sirola, Wafa Gull}
}

\usepackage{listings}
\lstdefinestyle{mesastyle}{basicstyle=\ttfamily\small,breaklines=true,frame=single,columns=fullflexible,showstringspaces=false}

\graphicspath{{./}{figures/}{figures/paper5/}{stageB/figures/}{stageB/tables/}}

\newcommand{\safeincludegraphics}[2][]{%
  \IfFileExists{#2}{\includegraphics[#1]{#2}}{%
    \fbox{\begin{minipage}[c][0.24\textheight][c]{0.88\linewidth}
      \centering\vspace{0.5em}\textbf{Missing figure file}\par\vspace{0.5em}
      \texttt{\detokenize{#2}}\par\vspace{0.5em}
      Place this figure file at the listed path or in the manuscript folder and recompile.
    \end{minipage}}}}

\newcommand{\safetableinput}[1]{%
  \IfFileExists{#1}{\input{#1}}{%
    \fbox{\begin{minipage}{0.92\linewidth}\centering\vspace{0.8em}\textbf{Missing table file}\par\vspace{0.5em}\texttt{\detokenize{#1}}\par\vspace{0.8em}\end{minipage}}}}

\newcommand{\dcep}{\ensuremath{\delta} Cephei}
\newcommand{\Msun}{\ensuremath{M_{\odot}}}
\newcommand{\Rsun}{\ensuremath{R_{\odot}}}

\newcommand{\msun}{\Msun}
\newcommand{\rsun}{\Rsun}

\newcommand{\MESA}{\textsc{MESA}}
\newcommand{\RSP}{\textsc{RSP}}
\newcommand{\MESARSP}{\MESA-\RSP}

\title{Evolutionary Period-Change Modeling of Delta Cephei with MESA Tracks}

\author[1,2,*]{Zuhoor Elahi}
\author[1]{Christopher Sirola}
\author[1]{Wafa Gull}

\affil[1]{Department of Physics and Astronomy, University of Southern Mississippi, Hattiesburg, MS, USA}
\affil[2]{Department of Physics, University of Karachi, Karachi, Pakistan}
\affil[*]{Corresponding author: zuhoor.elahi@usm.edu}

\date{}

\begin{document}
\maketitle

\begin{abstract}
Cepheid period changes provide a direct evolutionary diagnostic because the pulsation period responds to changes in stellar radius as a star crosses the instability strip.  We present a controlled \MESA{} evolutionary-track analysis for \dcep{} using nonrotating, no-wind models.  The period is estimated from the period--mean-density relation, \(P=Q[(R/R_\odot)^3/(M/M_\odot)]^{1/2}\), with \(Q=0.033~{\rm d}\).  The adopted comparison values are \(P_{\rm obs}=5.366531~{\rm d}\) and \(\dot P_{\rm obs}=-0.1006~{\rm s~yr^{-1}}\).  A solar-metallicity mass/overshoot grid identifies clean blueward negative-\(\dot P\) solutions near the observed period, but the best such case has \(\dot P=-0.6813~{\rm s~yr^{-1}}\), a factor of 6.77 too large in magnitude.  A metallicity pilot grid improves the result substantially, with a best clean-blueward model at \(M=5.90~\msun\), \(Z=0.012\), and \(f_{\rm ov,core}=0.010\), giving \(P=5.370677~{\rm d}\) and \(\dot P=-0.2460~{\rm s~yr^{-1}}\).  A local refinement around this solution does not improve beyond the same model.  Thus, metallicity refinement reduces the period-change mismatch by a factor of about 2.77 relative to the solar-metallicity baseline, but the final nonrotating no-wind fixed-\(Q\) model still overpredicts the observed magnitude of \(\dot P\) by a factor of about 2.45.  The result supports a blueward evolutionary interpretation while identifying rotation, mass loss, binary-related effects, and structure-dependent pulsation periods as the natural scope of a follow-up study.
\end{abstract}


\section{Introduction}
\label{sec:introduction}

Classical Cepheids are radially pulsating evolved stars whose periods are strongly connected to their mean densities.  They also underpin the classical period--luminosity relation first established from Magellanic Cloud variables \citep{leavitt1912}.  As a Cepheid evolves across the instability strip, its radius changes and the pulsation period changes secularly.  The sign of the period-change rate, \(\dot P\), is especially useful: blueward evolution generally produces decreasing radius and negative \(\dot P\), whereas redward evolution generally produces increasing radius and positive \(\dot P\).

The prototype classical Cepheid \dcep{} is a natural target for this test because it is well observed and has a period of about 5.37~d.  In this work we adopt
\begin{equation}
P_{\rm obs}=5.366531~{\rm d},
\label{eq:pobs}
\end{equation}
and use a working period-change constraint
\begin{equation}
\dot P_{\rm obs}=-0.1006~{\rm s~yr^{-1}}.
\label{eq:pdotobs}
\end{equation}
The negative sign of Eq.~\eqref{eq:pdotobs} motivates a search for a second-crossing, blueward evolutionary solution.  The central question of this paper is whether a controlled set of evolutionary tracks computed with Modules for Experiments in Stellar Astrophysics (\MESA) can reproduce both the observed period and the sign and approximate magnitude of the secular period change \citep{paxton2011,paxton2013,paxton2015,paxton2018,paxton2019,jermyn2023}.

Observed Cepheid period changes and their interpretation as evolutionary diagnostics have been studied extensively \citep[e.g.,][]{turner2006,engle2014,csornyei2022}.  Theoretical work has also examined how mass loss, rotation, metallicity, overshooting, and numerical choices affect Cepheid evolutionary tracks and period changes \citep{neilson2012,anderson2015,anderson2016,espinozaarancibia2022,ziolkowska2024,ziolkowska2026,smolec2026}.  The present contribution is a target-specific refinement study for \dcep{}.  It demonstrates that a fixed-\(Q\) period match can be misleading if the crossing direction and period-change sign are ignored, and it quantifies how mass, core overshooting, and metallicity affect the best nonrotating no-wind evolutionary solution.

This paper forms part of a broader dissertation program using \MESA{} and the \MESA{} Radial Stellar Pulsation module (\MESARSP{}) to model \dcep{}. The nonlinear pulsation workflow, period calibration, amplitude-control experiments, and model-acceptance criteria are presented by \citet{elahi2026nonlinear}, while the transformation of the resulting pulsation outputs into synthetic observed-band light curves is presented by \citet{elahi2026synthetic}. Those companion studies use fixed pulsation-model configurations, whereas the present paper follows full evolutionary tracks and examines the secular period change. The numerical parameters and results should therefore be regarded as complementary constraints rather than as alternative solutions for one identical stellar model.

\section{Method}
\label{sec:method}

\subsection{Period--mean-density estimate}
\label{subsec:pmeanrho}

The first-pass pulsation period is estimated using
\begin{equation}
P(t)=Q\left[\frac{(R(t)/R_\odot)^3}{M(t)/M_\odot}\right]^{1/2},
\label{eq:period_relation}
\end{equation}
where \(R(t)\) and \(M(t)\) are obtained from the \MESA{} evolutionary track.  We adopt a fixed fundamental-mode Cepheid-like value
\begin{equation}
Q=0.033~{\rm d}.
\label{eq:qadopted}
\end{equation}
This fixed-\(Q\) assumption provides a first-order evolutionary estimate.  The required value
\begin{equation}
Q_{\rm req}=P_{\rm obs}\left[\frac{M/M_\odot}{(R/R_\odot)^3}\right]^{1/2}
\end{equation}
is also recorded at the period-matching point as a consistency diagnostic.  A more complete treatment should calibrate \(Q\) using linear or nonlinear radial pulsation calculations along the evolutionary track.

Taking the logarithmic derivative of Eq.~\eqref{eq:period_relation} gives
\begin{equation}
\frac{\dot P}{P}=\frac{\dot Q}{Q}+\frac{3}{2}\frac{\dot R}{R}-\frac{1}{2}\frac{\dot M}{M}.
\label{eq:pdot_general}
\end{equation}
For the nonrotating, no-wind grids analyzed here, and under the fixed-\(Q\) approximation, the dominant term is expected to be the radius term,
\begin{equation}
\frac{\dot P}{P}\simeq \frac{3}{2}\frac{\dot R}{R}.
\label{eq:pdot_radius}
\end{equation}
Thus, crossing direction is central to the interpretation of \(\dot P\).

\subsection{Model grids}
\label{subsec:model_grid}

All models in the final calibration sequence are nonrotating and use no explicit wind mass loss.  The helium abundance is fixed at \(Y=0.270\), and the mixing-length theory (MLT) parameter is fixed at \(\alpha_{\rm MLT}=1.8\).  Core overshooting is represented with exponential overshooting at the top of the hydrogen-burning core.  The analysis was organized in two levels.  Stage A was a discovery grid at fixed composition, used to establish that period agreement alone is not a sufficient evolutionary criterion.  The Stage-A figures and tables are retained in Appendix~\ref{app:stageA}.  Stage B is the final refinement sequence used for the main quantitative result.  It contains three sub-stages, summarized in Table~\ref{tab:grid_summary}: B1 varies initial mass and core overshooting at solar-like metallicity, B2A tests metallicity near the best B1 region, and B2B performs a local refinement around the best B2A candidate.
All Stage-B calculations use Opacity Project at Livermore (OPAL) opacity tables with the adopted \texttt{a09} abundance mixture.

\begin{table}[H]
\centering
\caption{Summary of the Stage-B evolutionary model grid. All models are nonrotating, no-wind tracks with \(Y=0.270\), \(\alpha_{\rm MLT}=1.8\), and the adopted OPAL/\texttt{a09} opacity configuration.}

\label{tab:grid_summary}
\small
\begin{tabularx}{\textwidth}{lYcc}
\toprule
Stage & Purpose and parameter range & Models & Finished \\
\midrule
B1 & Solar-metallicity baseline: \(M=5.70,5.80,5.90,6.00,6.10~\msun\); \(f_{\rm ov,core}=0.010,0.014,0.016,0.018,0.020\); \(Z=0.014\). & 25 & 25 \\
B2A & Metallicity pilot: \(M=5.70,5.90~\msun\); \(f_{\rm ov,core}=0.010,0.014\); \(Z=0.010,0.012,0.014,0.016,0.020\). & 20 & 20 \\
B2B & Local refinement around the best B2A solution: \(M=5.85,5.90,5.95~\msun\); \(f_{\rm ov,core}=0.008,0.010,0.012\); \(Z=0.0110,0.0115,0.0120,0.0125\). & 36 & 36 \\
\bottomrule
\end{tabularx}
\end{table}

The opacity configuration is held fixed throughout the present evolutionary grid. The sensitivity of nonlinear \MESARSP{} periods and amplitude-growth diagnostics to the choice among native \MESA{} opacity tables has been examined separately using a fixed pulsation model \citep{elahi2026opacity}. Because that study did not compare
evolutionary tracks, its opacity-dependent results are not used to rank the evolutionary models considered here.

\subsection{Instability-strip and crossing classification}
\label{subsec:strip_identification}

A simple temperature window is used to identify Cepheid-like portions of each evolutionary track:
\begin{equation}
3.70 \leq \log T_{\rm eff} \leq 3.82.
\label{eq:simple_strip}
\end{equation}
This window is a first-order crossing criterion.  A segment is classified as clean blueward only if \(\Delta\log T_{\rm eff}>0\) and \(\Delta\log R<0\).  A segment is classified as redward if \(\Delta\log T_{\rm eff}<0\) and \(\Delta\log R>0\).  Cases with mixed behavior, such as increasing temperature while expanding, are treated as ambiguous diagnostics and are not accepted as clean blueward candidates.

\subsection{Period-change calculation and candidate ranking}
\label{subsec:pdot_calc}

For each crossing segment, the model point closest to \(P_{\rm obs}\) is identified using Eq.~\eqref{eq:period_relation}.  A local linear fit is applied to a small window around that point, using thirteen points when available, and the slope is converted from days per year to seconds per year.  A candidate is considered Cepheid-relevant if it lies within \(|P-P_{\rm obs}|\leq0.20~{\rm d}\), has \(\log L/L_\odot>3.0\), and has \(R>40~\rsun\).  A clean candidate must then satisfy three physical criteria: blueward crossing, \(\dot P<0\), and proximity to the observed period.  The formal score used for ranking combines period residual and period-change residual,
\begin{equation}
S=100\,|P-P_{\rm obs}|+10\,\left|\frac{\dot P-\dot P_{\rm obs}}{\dot P_{\rm obs}}\right|,
\label{eq:score}
\end{equation}
with penalties applied to non-Cepheid-relevant or non-blueward solutions.  The score is used only for ranking; the final interpretation is based on the physical crossing criteria.

\section{Stage-A Discovery Baseline}
\label{sec:stageA_baseline}

Before constructing the final Stage-B grid, we first carried out a Stage-A discovery calculation using nonrotating, no-wind \MESA{} evolutionary tracks at fixed composition, \(Z=0.014\) and \(Y=0.270\), with a fixed core-overshoot value \(f_{\rm ov,core}=0.016\).  The purpose of this initial stage was not to provide the final calibrated model, but to identify which parts of the evolutionary grid could approach the observed period of \dcep{} and to test whether period agreement alone was a sufficient selection criterion.

The Stage-A analysis showed that several tracks can pass near \(P_{\rm obs}=5.366531\,{\rm d}\) under the fixed-\(Q\) period--mean-density approximation.  However, the local crossing direction and the sign of \(\dot P\) separated physically acceptable from misleading period matches.  In particular, some models that matched the period occurred on redward segments or produced the wrong period-change sign.  The only strict Stage-A case satisfying blueward evolution and negative \(\dot P\) near the observed period was the \(6.0\,M_\odot\) model, but its local value, \(\dot P\simeq -0.8506\,{\rm s\,yr^{-1}}\), was substantially larger in magnitude than the adopted observed value, \(\dot P_{\rm obs}=-0.1006\,{\rm s\,yr^{-1}}\).

Stage A therefore established the main physical lesson used to design Stage B: matching \(P_{\rm obs}\) is not enough.  The refined grid must also require the correct crossing direction, the correct sign of \(\dot P\), and a period-change magnitude comparable to the observed secular period decrease.  The detailed Stage-A diagnostic figures and tables are provided in Appendix~\ref{app:stageA}; the main text below focuses on the Stage-B refinement that produced the final quantitative result.

\section{Stage-B Results}
\label{sec:results}

\subsection{Best candidates from the refinement sequence}
\label{subsec:best_by_stage}

Table~\ref{tab:best_by_stage} lists the best clean-blueward Cepheid-relevant candidate from each refinement stage.  In the B1 solar-metallicity grid, the best model has \(M=5.70~\msun\), \(Z=0.014\), and \(f_{\rm ov,core}=0.010\).  It matches the observed period to within \(-0.00508~{\rm d}\) and has the correct negative sign of \(\dot P\), but the predicted value, \(-0.6813~{\rm s~yr^{-1}}\), is too large in magnitude by a factor of 6.77.

The B2A metallicity pilot substantially improves the result.  The best B2A candidate has \(M=5.90~\msun\), \(Z=0.012\), and \(f_{\rm ov,core}=0.010\).  It gives \(P=5.370677~{\rm d}\), only \(+0.004146~{\rm d}\) from \(P_{\rm obs}\), and \(\dot P=-0.2460~{\rm s~yr^{-1}}\).  This reduces the period-change mismatch from a factor of 6.77 to a factor of 2.45.  The B2B local refinement includes the same physical point and does not improve beyond it; nearby variations in mass, metallicity, or overshooting produce larger \(|\dot P|\) or poorer period agreement.

\begin{table}[H]
\centering
\caption{Best clean-blueward Cepheid-relevant candidates from each Stage-B refinement step.  The B2B refinement did not improve beyond the B2A best model.}
\label{tab:best_by_stage}
\scriptsize
\resizebox{\textwidth}{!}{%
\begin{tabular}{llrrrrrrrrr}
\toprule
Stage & Run ID & $M/M_\odot$ & $Z$ & $f_{\rm ov}$ & $P$ (d) & $P-P_{\rm obs}$ (d) & $\dot P$ (s yr$^{-1}$) & $|\dot P|/|\dot P_{\rm obs}|$ & $\log T_{\rm eff}$ & $\log L/L_\odot$ \\
\midrule
B1 & B1\_M570\_Z014\_Y270\_FOV010\_ROT00\_WIND00 & 5.70 & 0.0140 & 0.010 & 5.361451 & -0.005080 & -0.681303 & 6.772 & 3.72612 & 3.31078 \\
B2A & B2A\_M590\_Z012\_Y270\_FOV010\_ROT00\_WIND00 & 5.90 & 0.0120 & 0.010 & 5.370677 & +0.004146 & -0.246028 & 2.446 & 3.75044 & 3.41903 \\
B2B & B2B\_M590\_Z0120\_Y270\_FOV010\_ROT00\_WIND00 & 5.90 & 0.0120 & 0.010 & 5.370677 & +0.004146 & -0.246028 & 2.446 & 3.75044 & 3.41903 \\
\bottomrule
\end{tabular}}
\end{table}

\subsection{Top final clean-blueward candidates}
\label{subsec:top_candidates}

The top B2B candidates confirm that the local refinement did not reveal a better solution than the B2A point.  The second-best B2B model, \(M=5.95~\msun\), \(Z=0.0125\), and \(f_{\rm ov,core}=0.010\), matches the period slightly better but predicts \(\dot P=-0.4263~{\rm s~yr^{-1}}\), corresponding to a mismatch factor of 4.24.  Other nearby B2B models also remain farther from the observed period-change magnitude.  A compact list of the leading models is given in Table~\ref{tab:top_candidates}.

\begin{table}[H]
\centering
\caption{Top clean-blueward Cepheid-relevant candidates from the final B2B local-refinement grid.}
\label{tab:top_candidates}
\scriptsize
\resizebox{\textwidth}{!}{%
\begin{tabular}{lrrrrrrr}
\toprule
Run ID & $M/M_\odot$ & $Z$ & $f_{\rm ov}$ & $P$ (d) & $P-P_{\rm obs}$ (d) & $\dot P$ (s yr$^{-1}$) & $|\dot P|/|\dot P_{\rm obs}|$ \\
\midrule
B2B\_M590\_Z0120\_Y270\_FOV010\_ROT00\_WIND00 & 5.90 & 0.0120 & 0.010 & 5.370677 & +0.004146 & -0.246028 & 2.446 \\
B2B\_M595\_Z0125\_Y270\_FOV010\_ROT00\_WIND00 & 5.95 & 0.0125 & 0.010 & 5.364154 & -0.002377 & -0.426349 & 4.238 \\
B2B\_M590\_Z0120\_Y270\_FOV012\_ROT00\_WIND00 & 5.90 & 0.0120 & 0.012 & 5.355846 & -0.010685 & -0.446846 & 4.442 \\
B2B\_M595\_Z0125\_Y270\_FOV012\_ROT00\_WIND00 & 5.95 & 0.0125 & 0.012 & 5.356029 & -0.010502 & -0.552269 & 5.490 \\
B2B\_M585\_Z0125\_Y270\_FOV012\_ROT00\_WIND00 & 5.85 & 0.0125 & 0.012 & 5.351129 & -0.015402 & -0.564470 & 5.611 \\
B2B\_M590\_Z0125\_Y270\_FOV012\_ROT00\_WIND00 & 5.90 & 0.0125 & 0.012 & 5.386359 & +0.019828 & -0.593403 & 5.899 \\
\bottomrule
\end{tabular}}
\end{table}

\subsection{Diagnostic figures}
\label{subsec:diagnostic_figures}

\begin{figure}[H]
\centering
\includegraphics[width=0.92\textwidth]{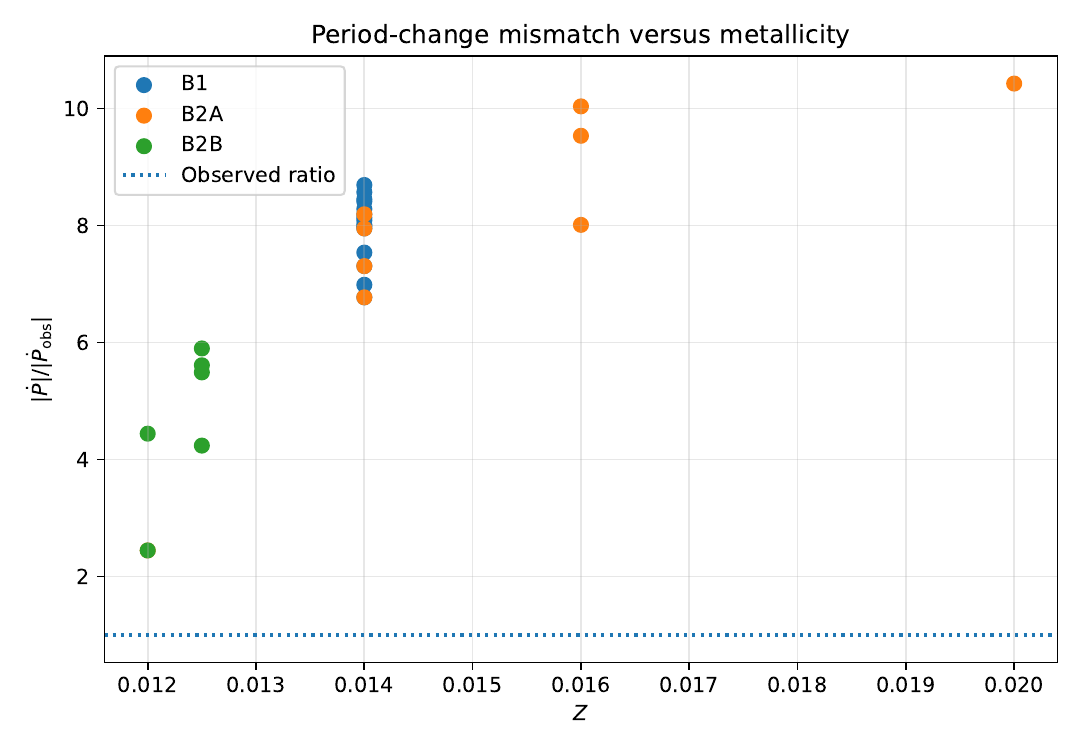}
\caption{Absolute period-change mismatch, \(|\dot P|/|\dot P_{\rm obs}|\), as a function of metallicity for clean-blueward Stage-B candidates.  The lowest mismatch occurs near \(Z=0.012\) in the tested grid.}
\label{fig:stageB_ratio_vs_Z}
\end{figure}

Figure~\ref{fig:stageB_pdot_vs_period} shows the clean-blueward candidates in the \(P\)--\(\dot P\) plane.  The best models lie close to the observed period but remain above the observed value in absolute period-change magnitude.  Figure~\ref{fig:stageB_ratio_vs_Z} shows the strong improvement near \(Z=0.012\), while higher metallicity tends to increase the magnitude of the predicted period change in the tested grid.

\begin{figure}[H]
\centering
\safeincludegraphics[width=0.92\textwidth]{stageB/figures/fig_stageB_pdot_vs_period.pdf}
\caption{Clean-blueward Stage-B candidates in the period--period-change plane.  The vertical dashed line marks \(P_{\rm obs}\), and the horizontal reference marks \(\dot P_{\rm obs}\).  The final best model matches the period closely but still predicts a period-change magnitude larger than observed.}
\label{fig:stageB_pdot_vs_period}
\end{figure}

\begin{figure}[H]
\centering
\includegraphics[width=0.80\textwidth]{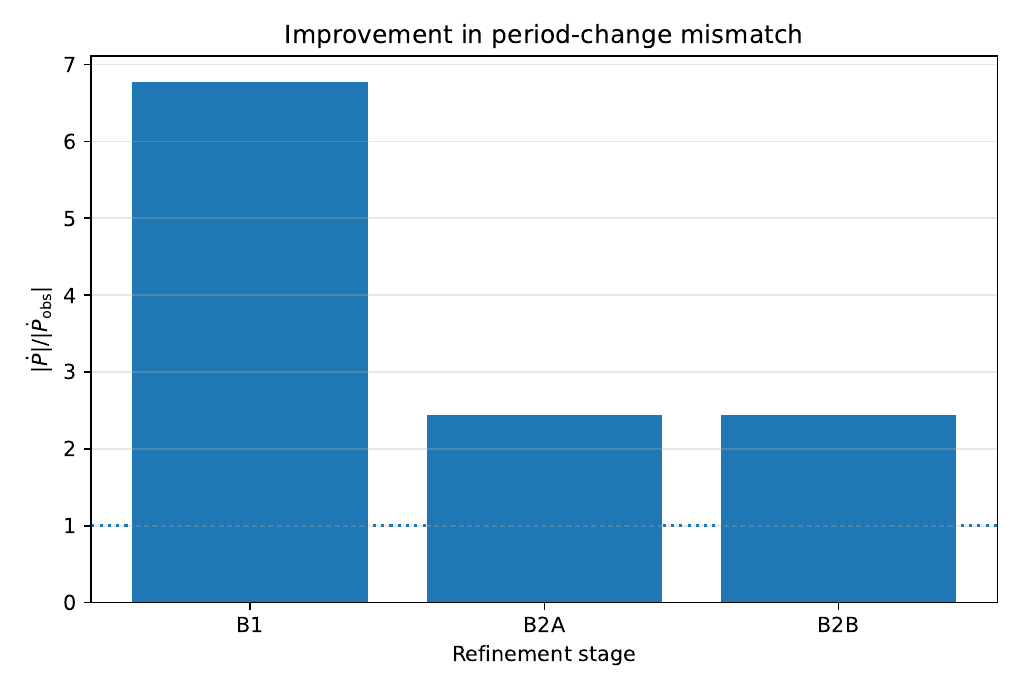}
\caption{Best-by-stage period-change mismatch.  The B2A metallicity pilot reduces the mismatch by a factor of about 2.77 relative to the B1 solar-metallicity baseline.  The B2B local refinement does not improve beyond the same best physical point.}
\label{fig:stageB_best_by_stage}
\end{figure}

\begin{figure}[H]
\centering
\includegraphics[width=0.92\textwidth]{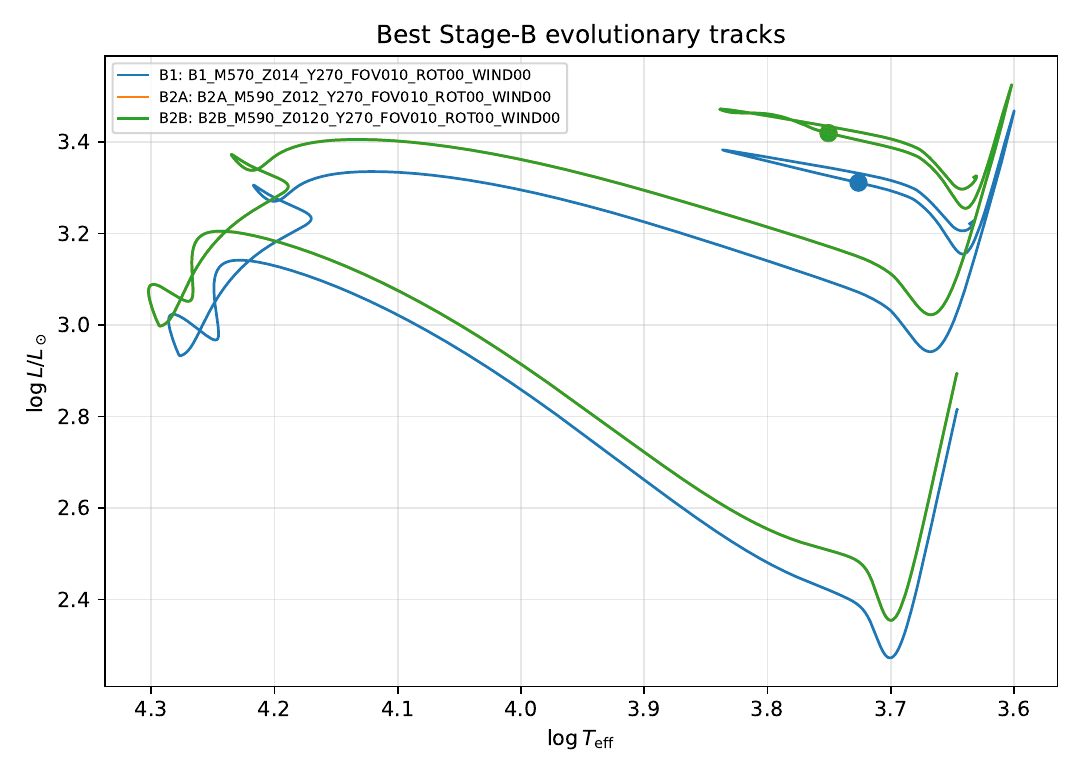}
\caption{HR diagram for the best-by-stage evolutionary tracks.  Markers indicate the local period-matching points used for the period-change comparison.  Hotter temperatures are plotted to the left.}
\label{fig:stageB_hr}
\end{figure}

Figure~\ref{fig:stageB_best_by_stage} summarizes the improvement from the solar-metallicity B1 baseline to the lower-metallicity B2A/B2B best solution.  Figure~\ref{fig:stageB_hr} shows the Hertzsprung--Russell (HR) tracks for the best-by-stage candidates and marks the period-matching points used in Table~\ref{tab:best_by_stage}.

\section{Discussion}
\label{sec:discussion}

\subsection{What the refinement sequence establishes}
\label{subsec:what_refinement_establishes}

The Stage-B sequence establishes three main points.  First, nonrotating no-wind tracks can reproduce the observed period and the correct negative sign of the period change when the crossing direction is treated carefully.  Second, the solar-metallicity mass/overshoot grid predicts a period-change magnitude that is too large by a factor of about 6.77 for the best clean-blueward candidate.  Third, a modest reduction in metallicity to \(Z=0.012\) substantially slows the predicted period evolution in the best model, reducing the mismatch to a factor of about 2.45.

The local B2B refinement provides an important negative result: simply adjusting mass by \(\pm0.05~\msun\), metallicity by small steps around \(Z=0.012\), or core overshooting around \(f_{\rm ov,core}=0.010\) does not produce a full match to \(\dot P_{\rm obs}\).  Therefore, the best candidate is not an isolated numerical accident, but the remaining mismatch appears robust within the tested nonrotating no-wind fixed-\(Q\) parameter space.

\subsection{Physical interpretation of the best model}
\label{subsec:best_model_interpretation}

The final best model has \(M=5.90~\msun\), \(Z=0.012\), \(Y=0.270\), and \(f_{\rm ov,core}=0.010\).  At the period-matching point it has \(P=5.370677~{\rm d}\), \(R=53.86~\rsun\), \(\log T_{\rm eff}=3.75044\), and \(\log L/L_\odot=3.41903\).  Its required pulsation constant is \(Q_{\rm req}=0.032975~{\rm d}\), very close to the adopted \(Q=0.033~{\rm d}\), so the period agreement is internally consistent within the fixed-\(Q\) approximation.

The remaining issue is not the period but the evolutionary speed.  The predicted value \(\dot P=-0.2460~{\rm s~yr^{-1}}\) has the correct sign but is still about 2.45 times too large in magnitude.  In the fixed-\(Q\), no-wind limit, this implies radius contraction through the instability strip that is still too rapid compared with the adopted observational constraint.

\subsection{Limitations and future physics}
\label{subsec:limitations}

Several limitations should be emphasized.  First, the instability strip is represented by a simple temperature window rather than luminosity-dependent theoretical blue and red edges.  Second, the pulsation constant is fixed at \(Q=0.033~{\rm d}\), although \(Q\) can vary with structure and evolutionary state.  Third, the grids intentionally do not include rotation, wind mass loss, or binary interaction.  These processes can change core mass, luminosity, blue-loop morphology, envelope structure, crossing timescale, and therefore \(\dot P\).  Fourth, the local period-change estimate depends on the time sampling and the exact segment classification.  These limitations do not invalidate the refinement result, but they define the next level of modeling.

The remaining discrepancy is therefore best interpreted as a controlled residual of the nonrotating, no-wind, fixed-\(Q\) framework rather than as a failure of the evolutionary interpretation.  The present paper isolates the baseline behavior produced by mass, core overshooting, and metallicity.  Rotation, mass loss, binary-related effects, and structure-dependent pulsation periods should be treated in a separate follow-up study.  That follow-up can begin from the best model found here, \(M=5.90\,\msun\), \(Z=0.012\), \(Y=0.270\), and \(f_{\rm ov,core}=0.010\), and test whether additional physics can reduce the period-change mismatch from \(|\dot P|/|\dot P_{\rm obs}|\simeq2.45\) toward unity without degrading the period match.

\section{Conclusions}
\label{sec:conclusions}

We constructed and analyzed a refined \MESA{} evolutionary grid for modeling the secular period change of \dcep{}.  The main conclusions are:

\begin{enumerate}
    \item The Stage-B nonrotating, no-wind grid contains 81 completed models: 25 in the B1 solar-metallicity mass/overshoot baseline, 20 in the B2A metallicity pilot, and 36 in the B2B local refinement.
    \item The fixed-\(Q=0.033~{\rm d}\) period--mean-density approximation can locate models that closely reproduce \(P_{\rm obs}=5.366531~{\rm d}\), but period agreement alone is not sufficient.  The crossing direction and \(\dot P\) sign are essential.
    \item The best B1 solar-metallicity clean-blueward candidate has \(M=5.70~\msun\), \(Z=0.014\), \(f_{\rm ov,core}=0.010\), \(P=5.361451~{\rm d}\), and \(\dot P=-0.6813~{\rm s~yr^{-1}}\).  Its period-change magnitude is too large by a factor of 6.77.
    \item The best B2A candidate has \(M=5.90~\msun\), \(Z=0.012\), \(f_{\rm ov,core}=0.010\), \(P=5.370677~{\rm d}\), and \(\dot P=-0.2460~{\rm s~yr^{-1}}\).  This reduces the period-change mismatch by a factor of about 2.77 relative to the B1 baseline.
    \item The B2B local refinement does not improve beyond the same physical model.  The final best candidate therefore remains the \(M=5.90~\msun\), \(Z=0.012\), \(f_{\rm ov,core}=0.010\) model.
    \item The final model reproduces the observed period closely and gives the correct negative period-change sign, but still overpredicts \(|\dot P|\) by a factor of about 2.45.  A fully quantitative match likely requires physics beyond the present nonrotating, no-wind, fixed-\(Q\) framework. Rotation, mass loss, binary-related effects, and structure-dependent pulsation periods are therefore reserved for a follow-up paper that starts from the best model found here.
\end{enumerate}

\appendix
\section{Stage-A Discovery-Grid Diagnostics}
\label{app:stageA}

Stage A was used as a discovery and baseline calculation before the refined Stage-B grid was constructed.  Its purpose was to test whether simple period agreement was sufficient for identifying an acceptable evolutionary model of \dcep{}.  The Stage-A models showed that several tracks can approach \(P_{\rm obs}\), but that period agreement alone does not determine the physical crossing.  Some period-matching cases occur on redward segments or have the wrong sign of \(\dot P\).  The Stage-A results therefore motivated the stricter Stage-B selection criteria used in the final analysis: proximity to \(P_{\rm obs}\), blueward evolution, negative \(\dot P\), and a period-change magnitude comparable to the observed value.

\begin{table}[H]
\centering
\caption{Stage-A discovery-grid setup.}
\label{tab:stageA_grid}
\scriptsize
\begin{tabular}{r r r r r r r l}
\toprule
$M_{\rm init}$ & $Z$ & $Y$ & $\alpha_{\rm MLT}$ & $f_{\rm ov,core}$ & Rotation & Wind & Status \\
$(M_\odot)$ & & & & & & & \\
\midrule
4.50 & 0.014 & 0.270 & 1.8 & 0.016 & 0 & 0 & completed \\
5.00 & 0.014 & 0.270 & 1.8 & 0.016 & 0 & 0 & completed \\
5.25 & 0.014 & 0.270 & 1.8 & 0.016 & 0 & 0 & completed \\
5.50 & 0.014 & 0.270 & 1.8 & 0.016 & 0 & 0 & completed \\
5.75 & 0.014 & 0.270 & 1.8 & 0.016 & 0 & 0 & completed \\
6.00 & 0.014 & 0.270 & 1.8 & 0.016 & 0 & 0 & completed \\
\bottomrule
\end{tabular}

\end{table}

\begin{table}[H]
\centering
\caption{Stage-A nearest-period summary. The table lists the model locations closest to \(P_{\rm obs}\) under the fixed-\(Q\) period--mean-density approximation.}
\label{tab:stageA_nearest_period}
\scriptsize
\resizebox{\textwidth}{!}{\begin{tabular}{r l r r r r r r}
\toprule
$M_{\rm init}$ & Run ID & Age & $P$ & $P-P_{\rm obs}$ & $\log L$ & $\log T_{\rm eff}$ & $R$ \\
$(M_\odot)$ & & (Myr) & (d) & (d) & & & $(R_\odot)$ \\
\midrule
4.50 & M45\_Z014\_OV016\_ROT00\_WIND00 & 131.795 & 5.35923 & -0.00730 & 2.892 & 3.6387 & 49.14 \\
5.00 & M50\_Z014\_OV016\_ROT00\_WIND00 & 109.902 & 5.36124 & -0.00529 & 3.155 & 3.6966 & 50.91 \\
5.25 & M525\_Z014\_OV016\_ROT00\_WIND00 & 90.990 & 5.35916 & -0.00737 & 2.974 & 3.6478 & 51.73 \\
5.50 & M55\_Z014\_OV016\_ROT00\_WIND00 & 81.627 & 5.37084 & +0.00431 & 3.005 & 3.6520 & 52.62 \\
5.75 & M575\_Z014\_OV016\_ROT00\_WIND00 & 73.686 & 5.36278 & -0.00375 & 3.047 & 3.6596 & 53.35 \\
6.00 & M60\_Z014\_OV016\_ROT00\_WIND00 & 66.895 & 5.36584 & -0.00070 & 3.207 & 3.6963 & 54.13 \\
\bottomrule
\end{tabular}
}
\end{table}

\begin{table}[H]
\centering
\caption{Stage-A local-fit period-change summary. Local fits near the period-matching points were used to estimate \(\dot P\).}
\label{tab:stageA_local_fit}
\scriptsize
\resizebox{\textwidth}{!}{\begin{tabular}{r l c l r r r r l}
\toprule
$M_{\rm init}$ & Run ID & Cross. & Direction & $P$ & $\dot P$ & $\log L$ & $\log T_{\rm eff}$ & Flag \\
$(M_\odot)$ & & & & (d) & (s yr$^{-1}$) & & & \\
\midrule
5.25 & M525\_Z014\_OV016\_ROT00\_WIND00 & 2 & blueward & 5.35555 & 0.0014 & 3.243 & 3.7152 & wrong\_pdot\_sign \\
5.50 & M55\_Z014\_OV016\_ROT00\_WIND00 & 3 & redward & 5.38294 & -0.0006 & 3.331 & 3.7331 & check \\
5.75 & M575\_Z014\_OV016\_ROT00\_WIND00 & 3 & redward & 5.37609 & -0.0014 & 3.412 & 3.7503 & check \\
6.00 & M60\_Z014\_OV016\_ROT00\_WIND00 & 2 & blueward & 5.35269 & -0.8506 & 3.489 & 3.7672 & good\_sign\_blueward \\
6.00 & M60\_Z014\_OV016\_ROT00\_WIND00 & 3 & redward & 5.40099 & 0.6185 & 3.504 & 3.7696 & wrong\_pdot\_sign \\
\bottomrule
\end{tabular}
}
\end{table}

\begin{table}[H]
\centering
\caption{Strict Stage-A blueward negative-\(\dot P\) candidates. These cases motivated the refined Stage-B search in mass, overshooting, and metallicity.}
\label{tab:stageA_strict_candidates}
\scriptsize
\resizebox{\textwidth}{!}{\begin{tabular}{l c l r r r r r r}
\toprule
Run ID & Cross. & Direction & $P$ & $\dot P$ & $\dot P-\dot P_{\rm obs}$ & $\log L$ & $\log T_{\rm eff}$ & $R$ \\
 & & & (d) & (s yr$^{-1}$) & (s yr$^{-1}$) & & & $(R_\odot)$ \\
\midrule
M60\_Z014\_OV016\_ROT00\_WIND00 & 2 & blueward & 5.35269 & -0.8506 & -0.7500 & 3.489 & 3.7672 & 54.05 \\
\bottomrule
\end{tabular}
}
\end{table}

\begin{figure}[H]
\centering
\includegraphics[width=0.92\textwidth]{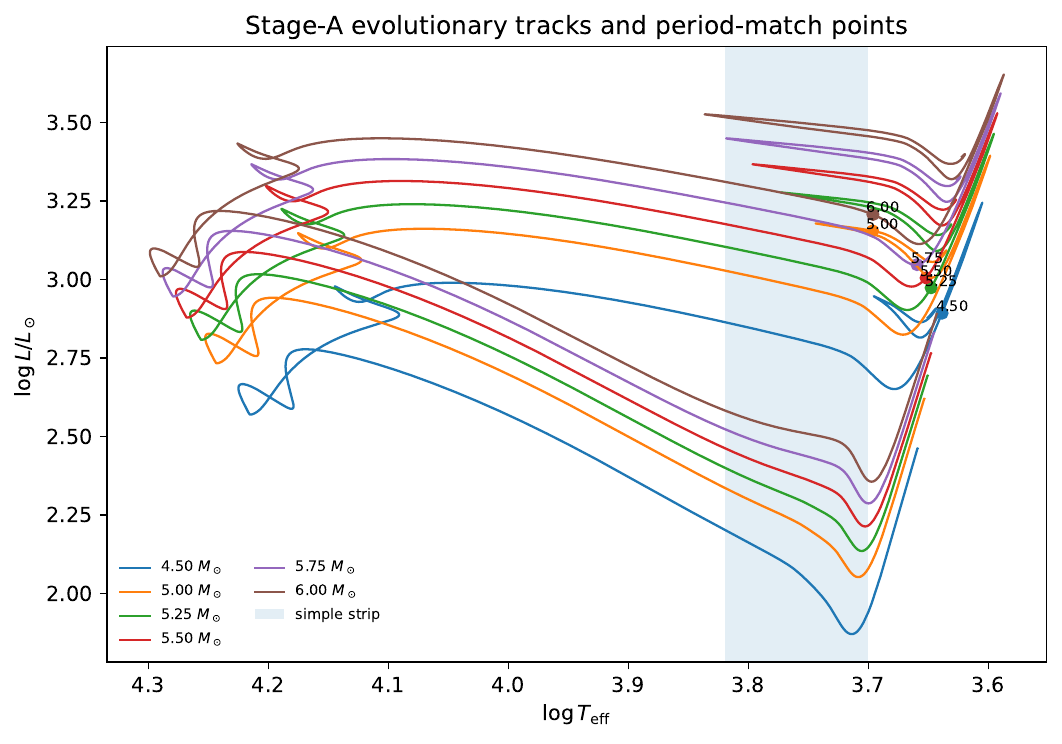}
\caption{Stage-A evolutionary tracks in the Hertzsprung--Russell diagram, shown relative to the instability-strip region used for the period-change search.}
\label{fig:stageA_hr_tracks}
\end{figure}

\begin{figure}[H]
\centering
\includegraphics[width=0.92\textwidth]{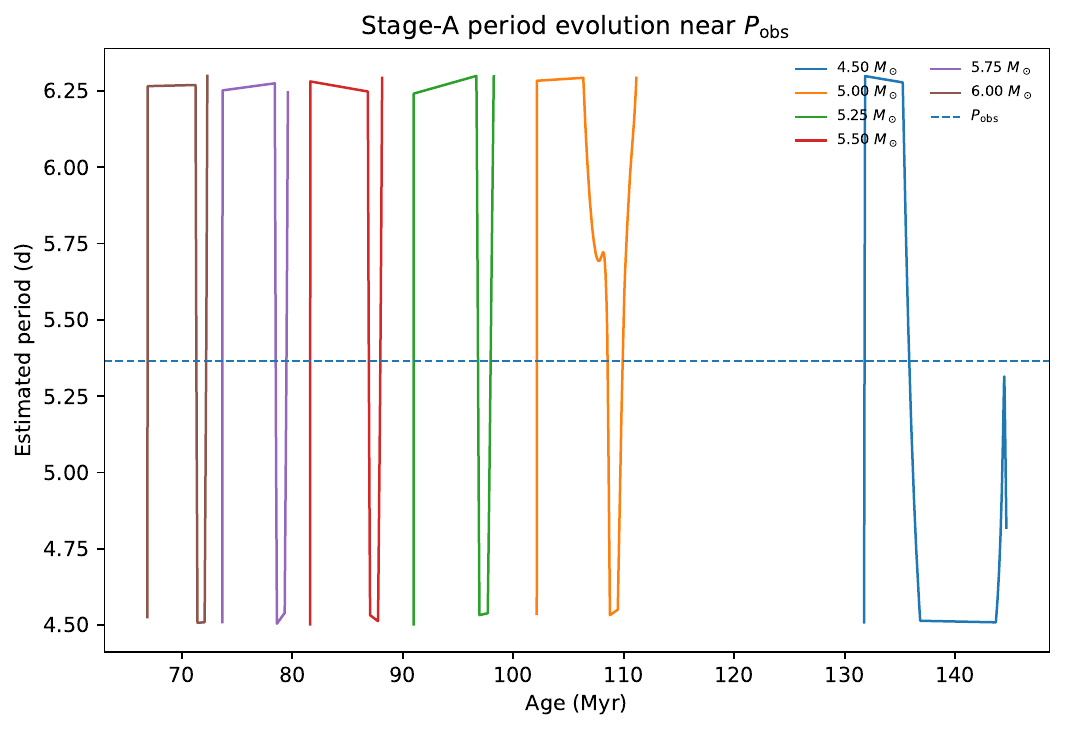}
\caption{Zoomed Stage-A view of model passages near the observed period of \dcep{}.  This figure illustrates why period agreement alone is not a sufficient physical selection criterion.}
\label{fig:stageA_period_zoom}
\end{figure}

\begin{figure}[H]
\centering
\includegraphics[width=0.92\textwidth]{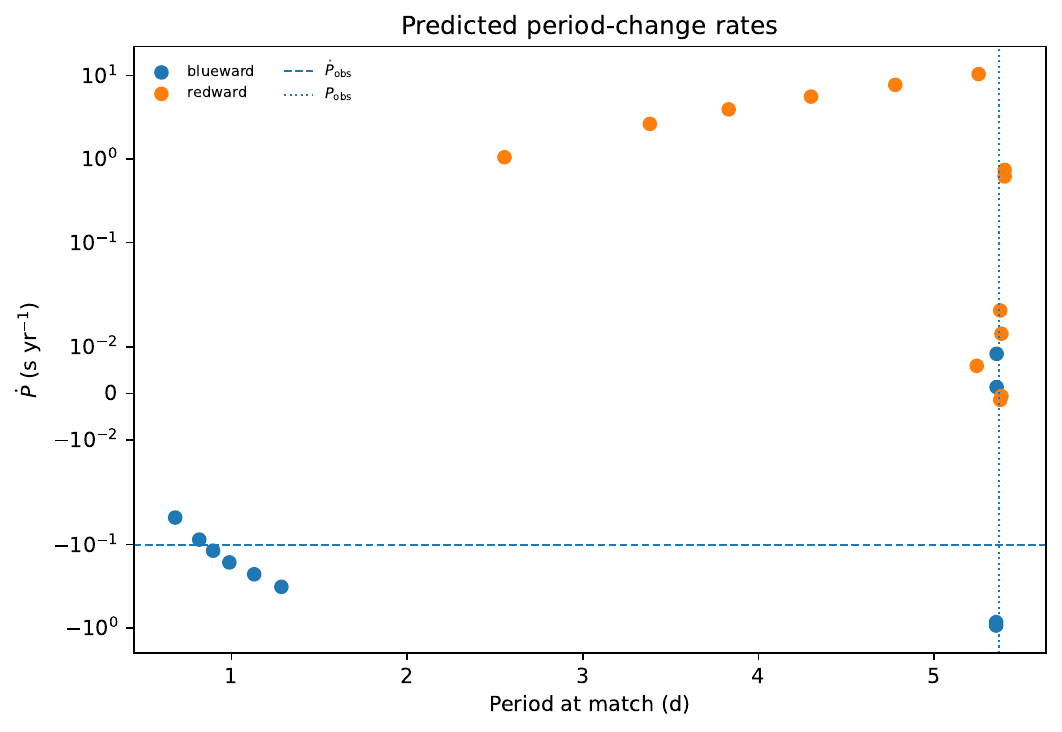}
\caption{Stage-A period-change diagnostic in the \(P\)--\(\dot P\) plane.  The observed period and observed period change define the two-dimensional comparison target.}
\label{fig:stageA_pdot_period}
\end{figure}

\begin{figure}[H]
\centering
\includegraphics[width=0.92\textwidth]{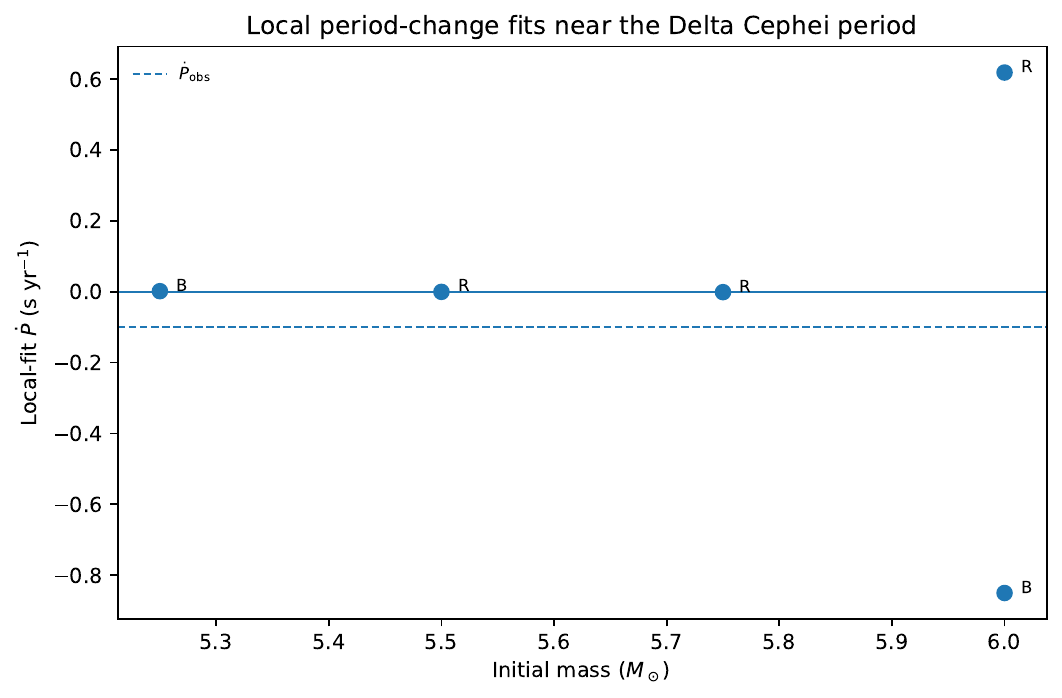}
\caption{Stage-A local-fit period-change summary.  Local fits near the period-matching points were used to estimate \(\dot P\) and to separate viable blueward negative-\(\dot P\) candidates from misleading period matches.}
\label{fig:stageA_local_fit_pdot}
\end{figure}

\begin{figure}[H]
\centering
\includegraphics[width=0.92\textwidth]{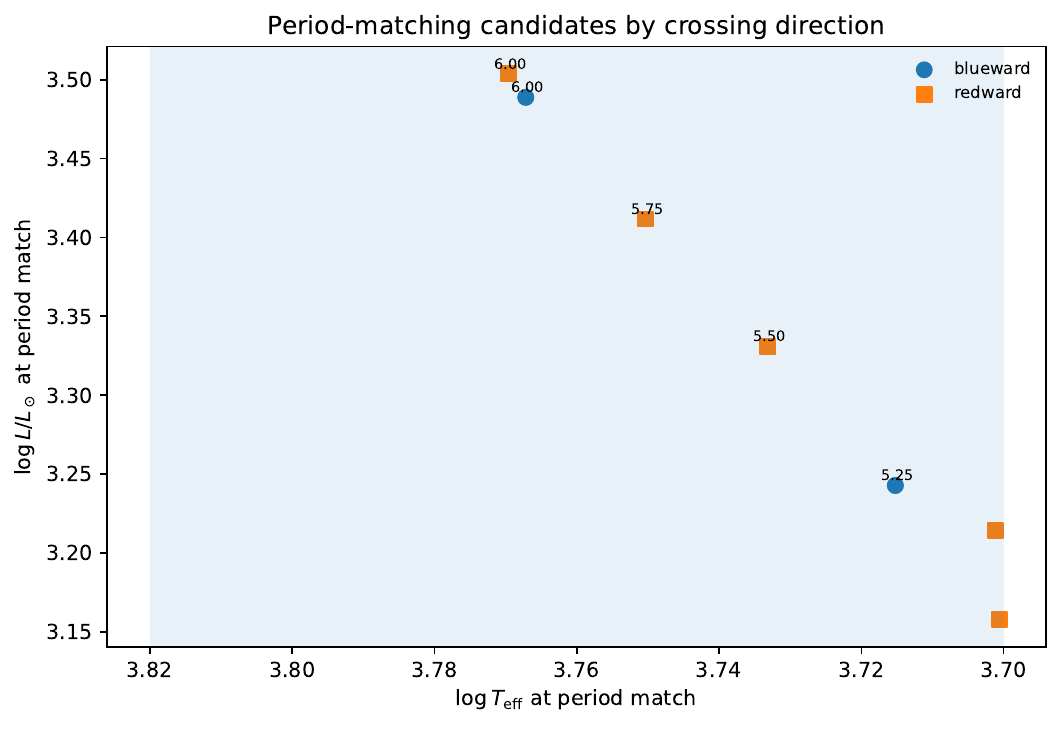}
\caption{Stage-A comparison of crossing behavior.  The diagnostic separates physically distinct evolutionary passages that may occur near similar pulsation periods.}
\label{fig:stageA_second_vs_third_crossing}
\end{figure}

\begin{figure}[H]
\centering
\includegraphics[width=0.92\textwidth]{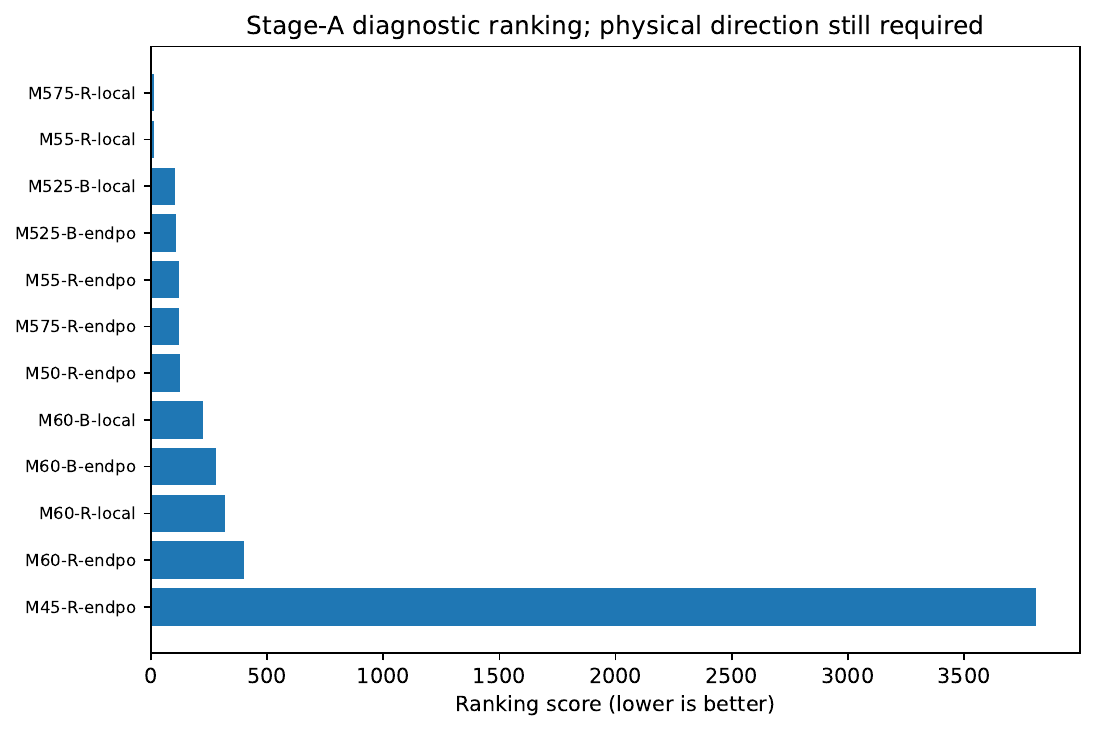}
\caption{Stage-A model-ranking map summarizing the discovery-grid outcome.  The ranking motivated the Stage-B refinement in mass, overshooting, and metallicity.}
\label{fig:stageA_model_ranking_map}
\end{figure}

\section*{Acknowledgments}

This work forms part of a dissertation project on \MESA{} and \MESARSP{} modeling of \dcep{}.  The authors acknowledge the use of \MESA{} and the supporting scientific Python ecosystem for model analysis and visualization.

\section*{Data and Code Availability}

The \MESA{} inlists, history files, analysis tables, and plotting scripts generated for this study are available from the corresponding author upon reasonable request.  The working analysis products include the Stage-A discovery-grid products, the final candidate ranking table, the best-by-stage table, and the Stage-B diagnostic figures described in the text.

\bibliographystyle{unsrtnat}
\bibliography{references}

@article{anderson2015,
  author  = {Anderson, R. I. and Saio, H. and Ekstr{\"o}m, S. and Georgy, C. and Meynet, G.},
  title   = {On the Effect of Rotation on Populations of Classical Cepheids. {I}. Predictions at Solar Metallicity},
  journal = {Astronomy \& Astrophysics},
  year    = {2015},
  volume  = {579},
  pages   = {A48}
}

@article{anderson2016,
  author  = {Anderson, R. I. and Saio, H. and Ekstr{\"o}m, S. and Georgy, C. and Meynet, G.},
  title   = {On the Effect of Rotation on Populations of Classical Cepheids. {II}. Pulsation Analysis for Metallicities 0.014, 0.006, and 0.002},
  journal = {Astronomy \& Astrophysics},
  year    = {2016},
  volume  = {591},
  pages   = {A8}
}

@article{csornyei2022,
  author  = {Cs{\"o}rnyei, G. and Szabados, L. and Moln{\'a}r, L. and others},
  title   = {Study of Changes in the Pulsation Period of 148 Galactic Classical Cepheids},
  journal = {Monthly Notices of the Royal Astronomical Society},
  year    = {2022},
  volume  = {511},
  pages   = {2125}
}

@article{engle2014,
  author  = {Engle, S. G. and Guinan, E. F. and Harper, G. M. and others},
  title   = {The Secret Lives of Cepheids: Evolutionary Changes and Pulsation-induced Heating of Classical Cepheid Atmospheres},
  journal = {The Astrophysical Journal},
  year    = {2014},
  volume  = {794},
  pages   = {80}
}

@article{espinozaarancibia2022,
  author  = {Espinoza-Arancibia, F. and Catelan, M. and Hajdu, G. and others},
  title   = {Period Change Rates of Large Magellanic Cloud Cepheids Using {MESA}},
  journal = {arXiv e-prints},
  year    = {2022},
  pages   = {arXiv:2209.10609},
  eprint  = {2209.10609},
  archivePrefix = {arXiv}
}

@article{jermyn2023,
  author  = {Jermyn, A. S. and Bauer, E. B. and Schwab, J. and others},
  title   = {Modules for Experiments in Stellar Astrophysics ({MESA}): Time-dependent Convection, Energy Conservation, Automatic Differentiation, and Infrastructure},
  journal = {The Astrophysical Journal Supplement Series},
  year    = {2023},
  volume  = {265},
  pages   = {15}
}

@article{leavitt1912,
  author  = {Leavitt, H. S. and Pickering, E. C.},
  title   = {Periods of 25 Variable Stars in the Small Magellanic Cloud},
  journal = {Harvard College Observatory Circular},
  year    = {1912},
  volume  = {173},
  pages   = {1}
}

@article{neilson2012,
  author  = {Neilson, H. R. and Cantiello, M. and Langer, N.},
  title   = {The Effects of Mass Loss on Cepheid Evolution and Pulsation},
  journal = {Astronomy \& Astrophysics},
  year    = {2012},
  volume  = {543},
  pages   = {A26}
}

@article{paxton2011,
  author  = {Paxton, B. and Bildsten, L. and Dotter, A. and others},
  title   = {Modules for Experiments in Stellar Astrophysics ({MESA})},
  journal = {The Astrophysical Journal Supplement Series},
  year    = {2011},
  volume  = {192},
  pages   = {3}
}

@article{paxton2013,
  author  = {Paxton, B. and Cantiello, M. and Arras, P. and others},
  title   = {Modules for Experiments in Stellar Astrophysics ({MESA}): Planets, Oscillations, Rotation, and Massive Stars},
  journal = {The Astrophysical Journal Supplement Series},
  year    = {2013},
  volume  = {208},
  pages   = {4}
}

@article{paxton2015,
  author  = {Paxton, B. and Marchant, P. and Schwab, J. and others},
  title   = {Modules for Experiments in Stellar Astrophysics ({MESA}): Binaries, Pulsations, and Explosions},
  journal = {The Astrophysical Journal Supplement Series},
  year    = {2015},
  volume  = {220},
  pages   = {15}
}

@article{paxton2018,
  author  = {Paxton, B. and Schwab, J. and Bauer, E. B. and others},
  title   = {Modules for Experiments in Stellar Astrophysics ({MESA}): Convective Boundaries, Element Diffusion, and Massive Star Explosions},
  journal = {The Astrophysical Journal Supplement Series},
  year    = {2018},
  volume  = {234},
  pages   = {34}
}

@article{paxton2019,
  author  = {Paxton, B. and Smolec, R. and Schwab, J. and others},
  title   = {Modules for Experiments in Stellar Astrophysics ({MESA}): Pulsating Variable Stars, Rotation, Convective Boundaries, and Energy Conservation},
  journal = {The Astrophysical Journal Supplement Series},
  year    = {2019},
  volume  = {243},
  pages   = {10}
}

@article{smolec2026,
  author  = {Smolec, R. and others},
  title   = {Toward a Comprehensive Grid of Cepheid Models with {MESA}. {III}. Evolutionary and Pulsation Relations},
  journal = {arXiv e-prints},
  year    = {2026},
  pages   = {arXiv:2603.26111},
  eprint  = {2603.26111},
  archivePrefix = {arXiv}
}

@article{turner2006,
  author  = {Turner, D. G. and Abdel-Sabour Abdel-Latif, M. and Berdnikov, L. N.},
  title   = {The Period Changes of Classical Cepheid Variables},
  journal = {Publications of the Astronomical Society of the Pacific},
  year    = {2006},
  volume  = {118},
  pages   = {410}
}

@article{ziolkowska2024,
  author  = {Zi{\'o}{\l}kowska, O. and Smolec, R. and Thoul, A. and Farrell, E. and Singh Rathour, R. and Hocd{\'e}, V.},
  title   = {Toward a Comprehensive Grid of Cepheid Models with {MESA}. {I}. Uncertainties of the Evolutionary Tracks of Intermediate-Mass Stars},
  journal = {The Astrophysical Journal Supplement Series},
  year    = {2024},
  volume  = {274},
  pages   = {30}
}

@article{ziolkowska2026,
  author  = {Zi{\'o}{\l}kowska, O. and Smolec, R. and Thoul, A. and Singh Rathour, R. and Hocd{\'e}, V.},
  title   = {Toward a Comprehensive Grid of Cepheid Models with {MESA}. {II}. Impact of Physical and Numerical Assumptions on Elemental Abundances},
  journal = {arXiv e-prints},
  year    = {2026},
  pages   = {arXiv:2602.08109},
  eprint  = {2602.08109},
  archivePrefix = {arXiv}
}

@misc{elahi2026synthetic,
  author        = {Elahi, Z. and Sirola, C. and Gull, W.},
  title         = {Synthetic Observed-Band Light Curves of Delta Cephei from MESA-RSP Models},
  year          = {2026},
  eprint        = {2607.10502},
  archivePrefix = {arXiv},
  primaryClass  = {astro-ph.SR},
  note          = {arXiv:2607.10502}
}

@misc{elahi2026nonlinear,
  author        = {Elahi, Z. and Sirola, C. and Gull, W.},
  title         = {Nonlinear MESA-RSP Modeling of Delta Cephei: Period Matching, Amplitude Control, and Model-Acceptance Diagnostics},
  year          = {2026},
  eprint        = {2607.10498},
  archivePrefix = {arXiv},
  primaryClass  = {astro-ph.SR},
  note          = {arXiv:2607.10498}
}

@misc{elahi2026opacity,
  author        = {Elahi, Z. and Sirola, C. and Gull, W.},
  title         = {Native-Opacity Sensitivity of a Fixed Delta Cephei MESA-RSP Pulsation Model},
  year          = {2026},
  eprint        = {2607.02439},
  archivePrefix = {arXiv},
  primaryClass  = {astro-ph.SR},
  note          = {arXiv:2607.02439}
}

\end{document}